# Possible thermal evolutionary pathways of irregular shaped small asteroids and planetesimals.

Sandeep Sahijpal, Department of Physics, Panjab University, Chandigarh 160014, India (sandeep@pu.ac.in)

**Abstract**

Two distinct thermal evolutionary pathways of irregular shaped small planetesimals in the early solar system have been studied. We have taken a case study of two S-type asteroids; (243) Ida and (951) Gaspra, on the basis of their precise physical dimensions accessed by the Philip Stooke small body 3-D shape models of the NASA Planetary Data System. The 3- D shape models for the two asteroids are based on the Galileo spacecraft fly-by mission. Based on our novel thermal evolutionary code for the precise shape of the asteroids we found that the small planetary bodies that accreted within the initial 2-3 million years (Myr) experienced sintering, whereas, the bodies formed afterwards were left unconsolidated, e.g., as a rubble pile. The former set of bodies could have formed by direct aggregation of nebula dust. Whereas, the majority of the rubble pile type small planetary bodies accreted latter by the assemblage of the fragmented debris of the initially existing planetesimals. These bodies cooled over tens of million years. Generations of small planetesimals formed over time in the early solar system evolved from an ensemble of compact consolidated bodies to rubble pile bodies due to collision induced fragmentation and re-accretion.

**Keyword:** Planetesimals; radioactive heating; sintering; thermal models; thermal metamorphism; Galileo spacecraft.

# 1. Introduction

The formation of the planetesimals in the early solar system commenced with the agglomeration of colliding dust grains in the solar nebula (Birnstiel et al., 2016; Weidenschilling, 2019). The gravitational collapse of an initially rotating presolar molecular cloud had earlier resulted in the formation of the protosun at the center around 4.56 billion years ago, surrounded by an accretion disc of gas and dust that is generally refereed as the solar nebula. The commencement of the formation of the planetesimals initiated with the growth of large dust accumulates due to collisional induced sticking of the fine dust particles through van der Waals interaction, at least within the inner accretion disk where ice condensation did not occur in a predominant manner (Birnstiel et al., 2016). The coagulation of dust grains resulted in the fractal growth of large accumulates with substantial porosity. The motion of the fine dust particles was initially coupled with the gas dynamics. Due to the growth in the size, the motion got decoupled. The growing particles eventually settled down towards the mid-plane and spiraled inwards within the accretion disk. The collisions experience by planetesimals, assisted by gravity, eventually resulted in the formation of the planets. Some of these earliest formed planetesimals survived as asteroids and trans-Neptunian objects (TNOs) by escaping accretion in any major planetary body, e.g., a planet or a satellite. These survived planetesimals provide significant information regarding the physico-chemical processes operating in the solar nebula and the earliest accreted small planetary bodies.

The planetesimals and the parent bodies of asteroids experienced a wide-range of planetary process in the early solar system (e.g., Dodd, 1981; Huss et al., 2006; Rudraswami et al. 2008). Depending upon their initial composition, especially, the ice to dust ratio, and the accretion timescales, the planetary processes range from thermal metamorphism, aqueous alteration to partial and extensive melting and planetary scale differentiation. While the impact induced heating became a prominent heat source for the thermal processing of large planetary bodies, e.g., the terrestrial planets, Moon, etc., the thermal processing of the majority of the planetesimals and asteroid parent bodies was brought about by the radiogenic decay energy of $^{26}$Al that is known to be present in the early solar system at a canonical initial level of $5\times10^{-5}$ for $^{26}$Al/$^{27}$Al, as deciphered from the earliest condensed Ca-Al-rich inclusion (MacPherson et al., 1995).

In order to understand the nature of the early thermal processing of the planetesimals, several thermal models based for the radiogenic heating have been proposed (Miyamoto et al., 1981; Bennett and McSween, 1996; Sahijpal, 1997; Ghosh and McSween Jr., 1998; Merk et al., 2002; Hevey and Sanders, 2006; Gupta and Sahijpal, 2010; Harrison and Grimm, 2010; Henke et al., 2012a, 2012b, 2016; Sramek et al. 2012; Monnereau et al., 2013; Golabek et al., 2014; Neumann et al., 2014, 2018; Gail et al., 2015; Bhatia and Sahijpal, 2017a, 2017b; Sahijpal and Goyal, 2018). All these models deal with the thermal evolution of spherical planetary bodies that are assumed to have acquired quasi-hydrostatic equilibrium. However, some of the planetesimals and parent bodies of asteroids are known to have not achieved such an equilibrium due to their small sizes and low gravity. An attempt was made recently by Sahijpal (2021) to develop 3-D (three dimensional) thermal models for the irregular shaped objects by assuming ellipsoidal shaped. In the present work, we have made a further attempt to develop thermal models of irregular shaped small objects of any arbitrary shape. The space-craft deduced 3-D shape and size models of small asteroids, TNOs and small satellites can be directly feed in our numerical code to deduce their thermal evolution.

In the present work, we have used the data set, '*EAR-A-5-DDR-STOOKE-SHAPE-MODELS-V2.0*' of the NASA Planetary Data System (PDS3) (Seidelmann et al., 2002; Archinal et al., 2011). This data set contains Philip Stooke small body 3-D shape models that are based on the optical photographs obtained from the NEAR, Galileo, Giotto, Vega 1, Vega 2 and Voyager missions. On the basis of the IAU recommendations for the Cartographic coordinates, the recommended 3-D shape models for the asteroid's surface is divided into longitudinal (0º − 360º) and latitudinal (-90º − 90º) angular grid, with an angular gap of 5º each, at specified radial distances of the surface from the body center.

The data in the case of the asteroids (243) Ida and (951) Gaspra is based on the Galileo spacecraft fly-by mission (D'Amario et al. 1992; Chapman 1996). The Galileo mission made the first-ever close observation of an asteroid, (951) Gaspra on 29$^{th}$ October, 1991, at a closest approach of ~1600 km (Belton et al. 1992). On 28$^{th}$ August, 1993, the spacecraft performed a fly-by at a distance of ~2400 km from the asteroid (243) Ida. It discovered its moon Dactyl. Ida has been classified as an S-type asteroid on the basis of its reflectance spectra (Wilson et al. 1999; Sullivan et al. 1996). It could have a density of 3.48 to 3.64 g cm$^{-3}$, with a porosity of 11–42%. It has been suggested that Ida could be a fragment of a large partially differentiated asteroid of 120 km diameter (Greenberg et al. 1996).

In the present work, we performed numerical simulations of the thermal evolution by considering two distinct S-type asteroids with shape and size comparable to (243) Ida and (951) Gaspra as a case study, without actually probing into the real origins of the two asteroids in their parent bodies. We are more focused towards the development of the numerical technique that can be used for a wide-range ensemble of small planetary bodies of different shapes and sizes. Our major goal is to understand the consequence of complete or partial consolidation of an initially accreted porous planetesimal on the long-term thermal evolution of the body. The details of the numerical technique are mentioned in the methodology section 2 along with the description of the various simulation parameters. The results obtained from the two set of thermal models, viz., the sets with and without thermal sintering (consolidation) of the initially accreted porous planetesimals are discussed in section 3. The major conclusions drawn from the work are finally summarize in section 4.

## 2. Methodology

The thermal models of the radiogenic heating of planetesimals and asteroids are based on the 3-D heat conduction partial differential equation (1) that involves a radioactive decay heat source, $^{26}$Al. Here, the temperature $T \equiv T(x, y, z, t)$ is estimated on the basis of the radioactive heat generation and conduction within planetesimal and asteroid. Q ($=2.2\times10^{-7}$ W kg$^{-1}$) is the initial heat production rate of the radioactive, $^{26}$Al, defined at the onset time of the formation of the Ca-Al-rich inclusions with a canonical value of $5\times10^{-5}$ for $^{26}$Al/$^{27}$Al. Here, we have assumed an H-chondrite value of 1.22 % for the stable $^{27}$Al (Dodd, 1981; Yomogida and Matsui, 1983). κ and c are the temperature dependent thermal diffusivity and specific heat, respectively. The equation (1) is sufficient enough to handle thermal metamorphism. The consolidation (sintering) of asteroids with an initially high porosity is incorporated appropriately by altering the spatial dimensions along with the variations in the thermal diffusivity.

$$\frac{\partial T}{\partial t} = \kappa \nabla^2 T + \frac{Q}{c} e^{-\lambda t} \quad (1)$$

The basic theoretical and numerical formulation adopted for solving the partial differential equation by using finite difference method is based on the recent numerical approach (Sahijpal, 2021). We solve the 3-D partial differential equation by converting it into finite difference equation (FDM) in 3-D (Lapidus and Pinder, 1982). We choose spatial (Δx, Δy, Δz) and temporal (Δt) grid elements in the FDM. In order to mathematically realize the exact shape and size of the asteroids on the basis of the space-craft deduced shape models, we develop a logic shape generator function that defines the physical dimension of the asteroids in the cartesian coordinate system. This function defines the physical existence of the asteroid in a binary manner, with the unity and zero values implying the existence and non-existence, respectively.

*2.1. Logic shape generator function*

The heat conduction partial differential equation (1) is solved in cartesian coordinates. We generate a hypothetical cuboid of adequate dimensions [2a×2b×2c] so that it can accommodate the size of the initially unconsolidated, instantaneously accreted asteroid in all three dimensions. The cuboid as well as the unconsolidated asteroid are assumed to be centered at the coordinates (a, b, c), with the origin of the cartesian coordinate system at (x, y, z) = (0, 0, 0), that represents one of the vertices of the cuboid. The initially accreted unconsolidated asteroid is assumed to be 1.2 times larger than the consolidated asteroid for which we have taken the shape model deduced by the fly-by mission. The distances in our present work are measured with respect to the origin of the coordinate system. All the spatial grid cartesian coordinates within the cuboid are initially converted into spherical polar coordinates (r, θ, φ) with respect to the center (a, b, c) of the cuboid. Here, θ ranges over 0º − 360º, and φ varies over 0º − 180º, and defines the azimuthal and polar angles, respectively. This facilitates a direct comparison between the *Spatial Grid* array, SG{r, θ, φ}, within the hypothetical cuboid in which the asteroid 3-D shape has to be mathematically embedded, with the actual shape models deciphered from the space-craft. As mentioned earlier, the 3-D shape models for the two asteroids are presented in spherical polar coordinates by dividing the asteroid's surface into longitudinal (0º − 360º) and latitudinal (-90º − 90º) angular grids, at an equal interval of 5º each, with specified radial distances of the surface from the body center (Seidelmann et al., 2002; Archinal et al., 2011). The data is in ASCII format, with three columns containing surface grid longitude, latitude and radial distance of the surface from the body center.

In order to implement these shape-size models into our numerical code, we initially transforms the latitudinal (-90º − 90º) angular coordinates to (180º − 0º) coordinate system, and create an *Asteroid-Surface Grid* array, ASG{$r_s$, θ, φ}. Here, '$r_s$' represents the distance of the unconsolidated asteroid's surface from the center (a, b, c) at particular azimuthal and polar angles.

The shape model of the asteroid is mathematically embedded into the hypothetical spatial grid cuboid [2a×2b×2c] by making a comparison of the radial distance, 'r' at every spatial grid from the center (a, b, c) within the cuboid as define by SG{r, θ, φ} with '$r_s$' as determines by ASG{$r_s$, θ, φ}. We use linear interpolation over the nearest neighboring φ angular grids for a specific value of θ, wherever the $r_s$ value is not available. A logic shape generator function is constructed in the cartesian coordinate system to define the 3-D physical shape and size of the asteroid by eq. (2). Here, the value of unity for G(x, y, z) represent the physical existence of asteroid. A value of zero represents asteroid's exterior region. It could also be mathematically possible to model the initial voids or porosities within the unconsolidated asteroid by zeros. Subsequent to consolidation, as these voids are removed, the logic shape generator function can take a unity values in these places. We have not attempted such a possible procedure here due to its complexity.

$$G(x, y, z) = 1, \quad \forall \quad r \leq r_s, \ r \in SG\{r, \theta, \varphi\}, \ r_s \in ASG\{r_s, \theta, \varphi\}$$

$$= 0, \quad \forall \quad r > r_s \quad (2)$$

*2.2. 3D-FDM*

The heat conduction partial difference eq. (1) is solved using finite difference method. Two distinct numerical approaches for the 3-D finite difference method have been recently developed for non-spherical bodies (Sahijpal, 2021). There is one explicit 3D-FDM method based on the direct conversion of the PDE into central difference equations involving the three spatial dimensions. The second approach involves semi-implicit Crank-Nicholson technique by using fractional-step method (FSM). Even though, the second approach is more efficient in terms of saving computational time, yet we have adopted the explicit 3D-FDM approach in the present work due to its numerical simplicity. We can numerically perform three orders of magnitude variations in thermal diffusivity during the consolidation (sintering) of the small bodies using the explicit approach in a much simpler manner. The eq. (1) can be converted into a series of finite difference eq. (3) across the entire physical extent of the asteroid by using an explicit central difference approximation. Here, the n+1$^{th}$ and n$^{th}$ thermal states are defined at two consecutive timesteps that are separated by a time Δt. The three consecutive spatial grids, i-1, i, i+1, are separated by a spatial gap of Δx along the x-axis. Identically, the three consecutive spatial grids, j-1, j, j+1, are separated by a spatial gap of Δy along the y-axis, and the three consecutive spatial grids, k-1, k, k+1, are separated by a spatial gap of Δz along the z-axis.

$$\frac{T_{i,j,k}^{n+1}-T_{i,j,k}^n}{\Delta t} \simeq \kappa \left[ \frac{T_{i-1,j,k}^n-2T_{i,j,k}^n+T_{i+1,j,k}^n}{\Delta x^2} + \frac{T_{i,j-1,k}^n-2T_{i,j,k}^n+T_{i,j+1,k}^n}{\Delta y^2} + \frac{T_{i,j,k-1}^n-2T_{i,j,k}^n+T_{i,j,k+1}^n}{\Delta z^2} \right] + \frac{Q}{c}e^{-\lambda t} \quad (3)$$

$$\frac{\kappa \Delta t}{\Delta x^2} = R_x; \quad \frac{\kappa \Delta t}{\Delta y^2} = R_y; \quad \frac{\kappa \Delta t}{\Delta z^2} = R_z \quad (4)$$

The dimensionless normalized thermal diffusivities are defined for the three dimensions by eq. (4). With an assumption of Δx = Δy = Δz, in all the performed simulations, the dimensionless normalized thermal diffusivity becomes identical in all directions. $R_x = R_y = R_z = R$.

$$T_{i,j,k}^{n+1} \simeq R(T_{i-1,j,k}^n + T_{i,j-1,k}^n + T_{i,j,k-1}^n) + (1-6R)T_{i,j,k}^n + R(T_{i+1,j,k}^n + T_{i,j+1,k}^n + T_{i,j,k+1}^n) + \frac{Q}{c}\Delta t\, e^{-\lambda t} \quad (5)$$

Thus, the temperature at any n+1$^{th}$ timestep can be deduced on the basis of temperature at the n$^{th}$ timestep for any spatial grid point using eq. (5).

## 2.3. Numerical implementation of sintering

Due to the fractal nature of the growth of large dust aggregates from fine dust in the solar nebula, the growth of the planetesimals resulted in substantial amount of porosity in the small planetary bodies. The random collisions experienced by the large rocky bodies during the accretion of planetesimals resulted in porosity as high as 50 % (Hevey and Sanders, 2006; Sahijpal et al., 2007; Henke et al., 2012a, 2012b, 2016; Sramek et al. 2012; Gail et al., 2015). These un-sintered (unconsolidated) planetary bodies have extremely low thermal diffusivity due to the presence of large number of voids that can transfer heat only by radiative heat transfer. Majority of these bodies are thus able to retain the radioactive decay heat, thereby, resulting in rapid heating. However, subsequent to the sintering (consolidation) experienced at high temperature ~700 K, the planetesimals experience compaction and loss of porosity. This results in an increase in the thermal diffusivity and, hence, a rapid cooling of the bodies. Some of the earlier works have indicated that there is an almost three orders of magnitude increase in the thermal diffusivity on account of compaction (Hevey and Sanders, 2006; Sahijpal et al., 2007; Henke et al., 2012a; Gail et al., 2015). In the case of small bodies, the compaction occurs due to heating around 650 – 700 K. However, on the basis of the observations of the vanishing pore space between the petrologic type 4 and type 5 H chondrites, the complete sintering might have occurred at higher temperature of ~970 K (Slater-Reynolds and McSween 2005; McSween et al. 1988).

We implemented the temperature dependence of thermal diffusivity for the unconsolidated as well as consolidated asteroid by following the eq. (6) for two distinct temperature regimes. A constant thermal diffusivity of the unconsolidated asteroid at temperature less than 670 K was assumed to be $6.4 \times 10^{-6}$ m$^2$ s$^{-1}$. The unconsolidated body was assumed to have uniform density with uniform distribution of voids. Using a three parameter Sigmoidal function, the thermal diffusivity was varied from this low value to higher values for a consolidated body at temperature $\geq$ 700 K (Yomogida and Matsui, 1983) over the temperature range of 670 – 700 K. This approach is distinct compared to the numerical approach adopted by Henke et al. (2012a). In spite of the differences, the essential change in the thermal diffusivity on account of compaction brings a rapid change in the heat conduction rate of the asteroid around 700 K by three orders of magnitude. The Sigmoidal function helps in a gradual thermal transition across three orders of magnitude variation in the thermal diffusivity without significant numerical instabilities (Sahijpal et al. 2007). Subsequent to the consolidation of the asteroid, the temperature dependence of the thermal diffusivity was adopted from the work by Yomogida and Matsui (1983).

$$\kappa(T) = 6.4 \times 10^{-6} + \frac{0.06762}{\left(1+e^{\frac{-(T_{i,j,k;t}-700)}{4.182}}\right)} \quad \forall \ T_{i,\,j,\,k;\,t} < 700 \text{ K} \quad (6)$$

$$= A + \frac{B}{T_{i,j,k;t}} \quad \forall \ T_{i,\,j,\,k;\,t} \geq 700 \text{ K},$$

A = 4.5×10$^{-7}$ m$^2$ s$^{-1}$ and B = 1.32 ×10$^{-4}$ m$^2$ K s$^{-1}$ (Yomogida and Matsui, 1983).

The rise in the thermal diffusivity on account of consolidation is because of the compaction of the asteroid due to the loss in porosity (Φ). We assumed a compaction of the unconsolidated body from the initial to the final state by a factor of 1.2 along the three cartesian coordinates in an identical manner, irrespective of the distinct shape. This corresponds to a net porosity loss (ΔΦ) of ~42 %. For every 10 K rise in the temperature of a specific spatial grid in the asteroid from 670 – 700 K, the porosity change (ΔΦ) was assumed to be ~11 %, ~12 % and ~18 %, respectively. We assume that the sintering occurs throughout the planetary body without any absence of regolith layer. The existence of regolith layer can substantially reduce the heat losses from the body.

The size reduction on account of compaction by a factor of 1.2 in the three orthogonal dimensions was performed by assuming that the rigidity of the entire body is not compromised. The shape of a basic spatial cube, defined by $\Delta x \times \Delta y \times \Delta z$, was retained along with its association with the neighboring cubes to maintain the integrity of the orthogonal spatial grid array. In actual practice, we could not visualize any mathematical procedure that could achieve this compaction in a gradual manner. It is much more complex to perform gradual compaction in 3-D compared to the 1-D compaction as performed in the earlier works (e.g., Sahijpal et al. 2007). It is not simply possible to perform compaction of 3-D body without compromising the cube structure of the unit spatial cell and the spatial grid array. However, since the compaction occurs in a very small temperature regime of 670 − 700 K, within a short duration of ~0.01 Myr., from an initially low thermal diffusivity to moderate thermal diffusivity, we can assume a rapid re-sizing of the orthogonal spatial grid from a pre-sintered to post-sintered state, at least in the scenarios with the onset time of planetesimal accretion ($T_{onset}$) considered in the present work.

*2.4. Numerical simulation parameters*

The solutions (eq. 5) of the finite difference equation are obtained by assuming a consolidated spatial grid interval, '$\Delta x$' (=$\Delta y$=$\Delta z$) of 0.1 km and a temporal grid interval, '$\Delta t$' of $5\times10^{-5} - 5\times10^{-4}$ Myr. in the simulations. The temperature dependence of the specific heat, 'c', in eq. (1) was adopted from the earlier works (Yomogida and Matsui, 1983; Sahijpal et al., 2007). It ranges from a value of 564 − 826 J kg$^{-1}$ over the relevant temperature range of 200 − 1200 K according to eq. (7) in the present work. A uniform density of 3560 kg m$^{-3}$ is assumed for the post-sintered asteroid. This amounts to a pre-sintered density of 1495 kg m$^{-3}$ for the asteroid with an initial porosity of 42%.

$$c(T_{i,j,k;t}) = 280 + 553 \times (1 - e^{-0.0036 \times T_{i,j,k;t}}) \qquad (7)$$

A constant temperature of 200 K was assumed across the entire un-sintered asteroid at the time of accretion. An identical constant surface temperature was also maintained throughout the entire thermal evolution of the asteroid. We have assumed an initial canonical $^{26}$Al/$^{27}$Al value of 5×10$^{-5}$ at the time of condensation of the earliest Ca-Al-rich inclusions (MaPherson et al., 1995). The instant accretion of the planetesimal (asteroid) was assumed to occur at a time, 'T$_{onset}$' subsequent to the condensation of the Ca-Al-rich inclusions with the canonical value. The onset time of the planetesimal accretion is the most critical parameter that determines its thermal evolution (e.g., Sahijpal, 2021). In the present work, we have simulated two scenarios corresponding to the T$_{onset}$ of 1.5 Myr. and 3.5 Myr. In the first case, the asteroid experiences a complete sintering, whereas, it the second scenario, the planetesimal remains as an unconsolidated body even after the thermal evolution. The cooling rates of the planetary bodies in these two scenarios are significantly different. The transition between the two scenarios occurs around T$_{onset}$ ~3 Myr.

## 3. Results and Discussion

Based on the 3-D shape models inferred for the asteroids (243) Ida and (951) Gaspra by the Galileo fly-by mission, we have attempted to develop the early thermal evolution of irregular shaped small asteroids having the shortest side physical dimensions of ≤ 20 km. The major emphasis is to establish the numerical techniques to understand the thermal evolution of small irregular shaped planetary bodies, e.g., asteroids, TNOs and small satellites, rather than focusing our objective specifically on the two asteroids. These specific asteroids could have been a part of a large planetary bodies that got fragmented subsequent to their thermal evolution (Greenberg et al. 1996). Among the 12 small planetary bodies for which the NASA Planetary Data System (PDS3) has provided the detailed shapes, three bodies are asteroids, namely, (243) Ida, (951) Gaspra and (253) Mathilde.

We deduced two distinct thermal evolutionary scenarios for the asteroid identical in shape and size with (951) Gaspra. These scenarios correspond to an accretion time, $T_{onset}$ of 1.5 Myr. and 3.5 Myr. The isothermal contours corresponding to these two scenarios at four distinct time-steps are presented in Figs. 1 & 4 for the three orthogonal dimensions, whereas, the thermal profiles at specific time intervals are plotted in the Figs. 2 & 5, respectively. The two scenarios exhibit distinct evolutionary trends. The consolidation of the body occurs in the scenario corresponding to $T_{onset}$ of 1.5 Myr. as a result of which the asteroid cools down rapidly due to three orders of magnitude enhancement in the thermal diffusivity subsequent to the sintering. In the second scenario corresponding to the $T_{onset}$ of 3.5 Myr., due to the delayed accretion, the radiogenic heating is not substantial enough to heat the body to achieve complete consolidation. Hence, the thermal diffusivity of the body stays low. As a result, the body subsequently experiences slow cooling over a time-scale of tens of million years (Fig. 4 & 5). The cooling rate becomes comparable to large planetary bodies (>100 km) (Sahijpal, 2021). It is possible that even in the absence of 'hot' pressing around 700 K, there could be 'cold' pressing of porous unconsolidated asteroid (see e.g., Henke et al. 2012a) on account of gravitational settling.

The thermal sintering in (951) Gaspra alike asteroid corresponding to $T_{onset}$ of 1.5 Myr. (Figs. 1−3) occurs around 0.24−0.25 Myr. subsequent to its instantaneous accretion. An identical compaction of the body in all the three dimensions is marked in the thermal profiles (Fig. 2) with respect to the origin of the coordinate system. The isothermal contours present the compaction in 3-D with respect to the body center (Fig. 1). These contours take the shape of the asteroid's planes. The size of the asteroid reduces considerably due to compaction. We started the simulation with an initially large unconsolidated body so as to explain the present final size of the asteroid. The sharp spikes in the shape of the modelled asteroid surface along the XY planes (Fig. 1), prominent in the unconsolidated body isotherms, are due to the limitations in our linear interpolation over the nearest neighboring φ angular grids for a specific value of θ while defining the shape of the model for simulation (eq. 2). Subsequent to the sintering, the body achieved a maximum temperature of ~770 K around 0.3 Myr. after the onset of accretion. The isothermal contours along eight distinct XY planes that are orthogonal to Z-axis are graphically presented in Fig. 3 corresponding to this time interval. In the assumed absence of regolith with low thermal diffusivity, the temperature of the post-sintered body drops rapidly to < 300 K within 2 Myr (Fig. 1 & 2). The rapid cooling occurs along the shorter Y- and Z-axis.

The pre-sintered thermal evolution of the (253) Ida alike asteroid with $T_{onset}$ of 1.5 Myr. (Figs. 6 & 7) is almost identical to that of (951) Gaspra alike asteroid with an identical accretion timescale. The size of the asteroid does not matter in determining the thermal evolution as long as the thermal diffusivity is extremely small during the un-sintered stage. The complete consolidation of the (253) Ida alike asteroid occurs around 0.25 Myr. after accretion (Fig. 7). Subsequently, the asteroid acquired high temperature comparable to even the melting point of metallic iron. It cooled down slowly over ~ 5 Myr. compared to the (951) Gaspra model (Fig. 2). The shortest Z-axis provides the fastest cooling path.

On the basis of the two distinct thermal evolutionary pathways deduced for the small planetary bodies, we could broadly divide the time regime for the accretion of small planetesimals in the early solar system if we assume a uniform distribution of $^{26}Al$ in the accreting region. These time regimes can be broadly demarcated by 2-3 Myr. in the early solar system depending upon the initial composition of the body. The small planetary bodies accreted during the initial 2-3 Myr. experienced thermal sintering, and hence subsequently cooled rapidly. These bodies could have formed either by the direct accretion of solar nebula dust (Weidenschilling, 2019), or by aggregation of fragmented debris of the initially formed planetesimals. The planetesimals (asteroids) formed after the initial 2-3 Myr. were mostly the result of accretion of the fragmented debris of the initially existing planetesimals. The accretion of these bodies through direct solar nebula dust seems to be less likely due to their delayed formation. These bodies experienced very slow cooling over tens of million years, and could have survived as rubble piles.

In general, the formation of the small irregular shaped planetesimals (with the shortest side physical dimension ≤ 20 km), through the accretion of solar nebula dust, initiated during the early phases of the solar system. These initially formed bodies experienced thermal sintering and rapid cooling. Some of the earliest formed bodies, with an accretion time-scale < 1 Myr., could have experienced widespread melting and planetary scale differentiation. The frequent collisions experienced by the planetesimals could have produced fragmentation debris. These debris from several parent bodies accreted further to form the successive generations of small planetesimals. The thermal induced sintering could have occurred in these successive generations of small bodies during the initial 2-3 Myr. However, the small planetesimals formed in the subsequent generations were left un-sintered as rubble pile assemblages.

## 4. Conclusions

We present numerical technique for understanding the thermal evolution of small irregular shaped planetesimals, with the shortest side physical dimension ≤ 20 km. The 3-D shape models deduced for (243) Ida and (951) Gaspra from the Galileo fly-by missions were mathematically modelled to understand the thermal evolution of planetesimals and asteroid with identical dimensions. We could achieve the state transition of the initially unconsolidated planetary body to a final sintered body by numerically modelling the thermal sintering in a parametric form. Two distinct thermal evolutionary pathways were deduced for the sample asteroids. The scenario with an early accretion of planetesimal within the initial 2-3 Myr. results in complete thermal sintering of the small bodies, whereas, the bodies formed afterwards were left un-sintered. In general, during the repeated cycles of the formation and collision induced fragmentation of several generations of small planetesimals, the solar system evolved from an initial ensemble of compact small objects to an ensemble of rubble piles.

## Author Information

## Affiliations

**Department of Physics, Panjab University, Chandigarh 160014, India**

Sandeep Sahijpal

## Contributions

S. Sahijpal developed the numerical technique for mathematically modelling the thermal evolution of the small irregular shape asteroids and planetesimals. He wrote the manuscript.

## Corresponding author

Correspondence to Sandeep Sahijpal

# References

Archinal B A, A'Hearn M F, Bowell E, Conrad A, Consolmagno G J, Courtin R, Fukushima T, Hestroffer D, Hilton J L, Krasinsky G A, Neumann G, Oberst J, Seidelmann P K, Stooke P, Tholen D J, Thomas P C and Williams I P 2011 Report of the IAU working group on Cartographic coordinates and rotational elements: 2009; *Celestial Mechanics and Dynamical Astronomy* **109** 101-135, https://doi.org/10.1007/s10569-010-9320-4

Belton M J S et al., 1992 Galileo encounter with 951 Gaspra: first pictures of an asteroid; *Science* **257** 1647–52, https://doi.org/10.1126/science.257.5077.1647

Bennett M E and McSween H Y 1996 Shock features in iron-nickel metal and troilite of L-group ordinary chondrites; *Meteor. Planet. Sci.* **31** 255–264, https://doi.org/10.1111/j.1945-5100.1996.tb02021.x

Bhatia G K and Sahijpal S 2017a Did $^{26}$Al and impact-induced heating differentiate mercury? *Meteor. Planet. Sci.* **52** 295–319, https://doi.org/10.1111/maps.12789.

Bhatia G K and Sahijpal S 2017b Thermal evolution of the Trans-Neptunian Objects (TNOs), icy satellites and minor icy planets in the early solar system; *Meteor. Planet. Sci.* **52** 2470–2490, https://doi.org/10.1111/maps.12952.

Birnstiel T, Fang M and Johansen A 2016 Dust evolution and the formation of planetesimals; *Space Sci. Rev.* **205** 41–75,

Chapman C R 1996 S-type asteroids, ordinary chondrites, and space weathering: The evidence from Galileo's fly-bys of Gaspra and Ida; *Meteoritics* **31** 699–725, https://doi.org/10.1111/j.1945-5100.1996.tb02107.x

D' Amario L A, Bright L E and Wolf A 1992 Galileo trajectory design; *Space Sci. Rev.* **60** 23–78, https://doi.org/10.1007/BF00216849

Dodd R T 1981 Meteorites: A petrologic-chemical synthesis, 1st ed. New York: Cambridge University Press. 368 p.

Gail H -P, Henke S and Trieloff M 2015 Thermal evolution and sintering of chondritic planetesimals. II. Improved treatment of the compaction process; *Astron. Astrophys.* **576** A60, https://doi.org/10.1051/0004-6361/201424278.

Ghosh A and McSween H Y Jr. 1998 A thermal model for the differentiation of Asteroid 4 Vesta based on radiogenic heating; *Icarus* **134** 187–206, https://doi.org/10.1006/icar.1998.5956

Golabek G J, Bourdon B and Gerya T V 2014 Numerical models of the thermomechanical evolution of planetesimals: Application to the acapulcoite lodranite parent body; *Meteor. Planet. Sci.* **49** 1083–1099, https://doi.org/10.1111/maps.12302.

Greenberg R, Bottke W F, Nolan M, Geissler P E, Petit J M, Durda D D, Asphaug E and Head J 1996 Collisional and Dynamical History of Ida; *Icarus* **120** 106–118, https://doi.org/10.1006/icar.1996.0040

Gupta G and Sahijpal S 2010 Differentiation of Vesta and the parent bodies of other achondrites; *J. Geophys. Res.* **115** E08001, https://doi.org/10.1029/2009JE003525.

Harrison K P and Grimm R E 2010 Thermal constraints on the early history of the H chondrite parent body reconsidered; *Geochim. Cosmochim. Acta* **74** 5410–5423, https://doi.org/10.1016/j.gca.2010.05.034

Henke S, Gail H -P, Trieloff M, Schwarz W H and Kleine T 2012a Thermal evolution and sintering of chondritic planetesimals; *Astron. Astrophys.* **537** A45, https://doi.org/10.1051/0004-6361/201117177.

Henke S, Gail H -P, Trieloff M and Schwarz W H 2012b Thermal evolution model for the H chondrite asteroid-instantaneous formation versus protracted accretion; *Icarus* **226** 212–228, https://doi.org/10.1051/0004-6361/201219100.

Henke S, Gail H -P, Trieloff M and Schwarz W H 2016 Thermal evolution and sintering of chondritic planetesimals. III. Modelling the heat conductivity of porous chondrite material; *Astron. Astrophys.* **589** A41, https://doi.org/10.1051/0004-6361/201527687.

Hevey P J and Sanders I S 2006 A model for planetesimal meltdown by $^{26}$Al and its implications for meteorite parent bodies; *Meteor. Planet. Sci.* **41** 95–106, https://doi.org/10.1111/j.1945-5100.2006.tb00195.x.

Huss G R, Rubin A E and Grossman J N 2006 Thermal Metamorphism in Chondrites*; Meteorites and the Early Solar System II*, Lauretta D S McSween Jr. H Y (Eds.), University of Arizona Press, Tucson, pp. 567–586.

Lapidus L and Pinder G F 1982 Numerical solution of partial differential equations in science and engineering. Wiley-Interscience Publications, New York, p. 677.

MacPherson G J, Davis A M and Zinner E K 1995 The distribution of aluminum-26 in the early solar system—A reappraisal; *Meteoritics* **30** 365–386, https://doi.org/10.1111/j.1945-5100.1995.tb01141.x

McSween H Y Jr, Sears D W G and Dodd R T 1988 Thermal metamorphism; *Meteorites and the Early Solar System*, Kerridge J F and Matthews M S (Eds.), University of Arizona Press, Tucson, pp. 102-113.

Merk R, Breuer D and Spohn T 2002 Numerical modeling of $^{26}$Al induced radioactive melting of asteroids considering accretion; *Icarus* **159** 183–191, .

Miyamoto M, Fujii N and Takeda H 1981 Ordinary chondrite parent body: An internal heating model; *Proc. Lunar Planet. Sci. Conf. Lett.* **12B**, 1145.

Monnereau M, Toplis M J, Baratoux D and Guignard J 2013 Thermal history of the H chondrite parent body: Implications for metamorphic grade and accretionary timescales; *Geochim. Cosmochim. Acta* **119** 302–321,

Neumann W, Breuer D and Spohn T 2014 Modelling of compaction in planetesimals; *Astron. Astrophys.* **567** A120, https://doi.org/10.1051/0004-6361/201423648.

Neumann W, Henke S, Breuer D, Gail H -P, Schwarz W H, Trieloff M, Hopp J and Spohn T 2018 Modeling the evolution of the parent body of acapulcoites and lodranites: a case study for partially differentiated asteroids; *Icarus* **311** 146–169, https://doi.org/10.1016/j.icarus.2018.03.024.

Rudraswami N G, Goswami J N, Chattopadhyay B, Sengupta S K and Thapliyal A P 2008 $^{26}$Al records in chondrules from unequilibrated ordinary chondrites: II. Duration of chondrule formation and parent body thermal metamorphism; *Earth and Planetary Science Letters* **274** 93-102, https://doi.org/10.1016/j.epsl.2008.07.004

Sahijpal S 1997 Isotopic studies of the early solar system objects in meteorites by an ion microprobe. Ph.D. thesis. PRL, India.

Sahijpal S 2021 Thermal evolution of non-spherical asteroids in the early solar system; *Icarus* **362** 114439, https://doi.org/10.1016/j.icarus.2021.114439.

Sahijpal S and Goyal V 2018 Thermal evolution of the early Moon; *Meteorit. Planet. Sci.* **53** 2193–2211, https://doi.org/10.1111/maps.13119.

Sahijpal S, Soni P and Gupta G 2007 Numerical simulations of the differentiation of accreting planetesimals with $^{26}$Al and $^{60}$Fe as the heat sources; *Meteor. Planet. Sci.* **42** 1529–1548, .

Seidelmann P K, Abalakin V K, Bursa M, Davies M E, de Bergh C, Lieske J H, Oberst J, Simon J L, Standish E M, Stooke P J and Thomas P C 2002 Report of the IAU/IAG working group on Cartographic coordinates and rotational elements of the planets and satellites: 2000; *Celestial Mechanics and Dynamical Astronomy* **82** 83-100, https://doi.org/10.1023/A:1013939327465

Sramek O, Milelli L, Ricard Y and Labrosse S 2012 Thermal evolution and differentiation of planetesimals and planetary embryos; *Icarus* **217** 339–354, https://doi.org/10.1016/j.icarus.2011.11.021

Slater-Reynolds V and McSween H Y Jr 2005 Peak metamorphic temperatures in type 6 ordinary chondrites:An evaluation of pyroxene and plagioclase geothermometry; *Meteor. Planet. Sci.* **40** 745–754, https://doi.org/10.1111/j.1945-5100.2005.tb00977.x

Sullivan R J, Greeley R, Pappalardo R, Asphaug E, Moore J M, Morrison D, Belton M J S, Carr M, et al. 1996 Geology of 243 Ida; *Icarus* **120** 119–139,

Yomogida K and Matsui T 1983 Physical properties of ordinary chondrites; *J. Geophys. Res.* **88** 9513–9533, https://doi.org/10.1029/JB088iB11p09513.

Wilson L, Keil K and Love S J 1999 The internal structures and densities of asteroids; *Meteorit. Planet. Sci.* **34** 479–483, https://doi.org/10.1111/j.1945-5100.1999.tb01355.x

Weidenschilling S J 2019 Accretion of the asteroids: Implications for their thermal evolution; *Meteorit. Planet. Sci.* **54** 1115-1132, https://doi.org/10.1111/maps.13270

**Figures**

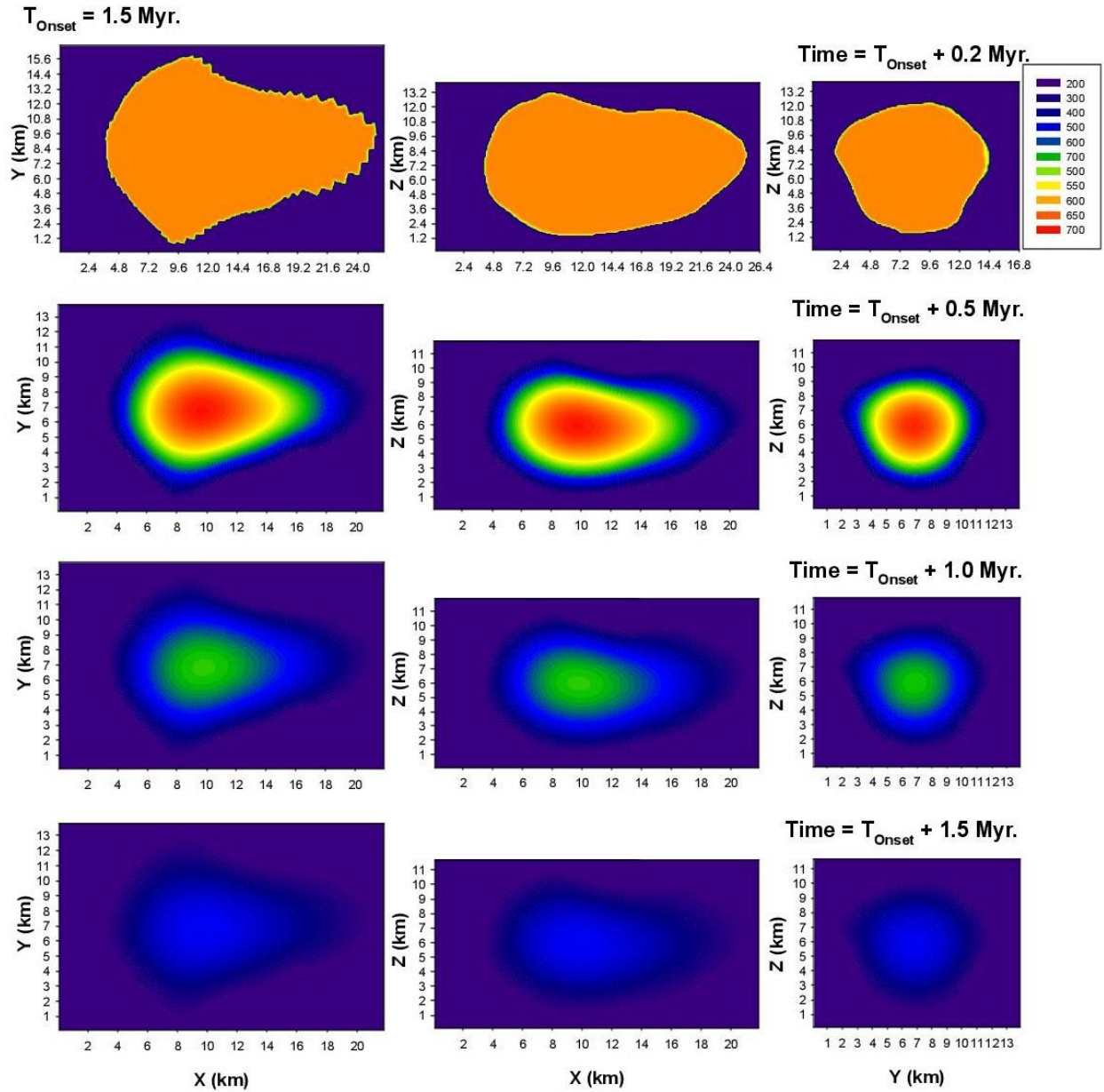

**Figure 1.** Temperature (K) contours of (951) Gaspra alike asteroid with $T_{Onset}$ = 1.5 Myr. Temperature contours for the three planes, XY, XZ and YZ through the center of the body at four distinct time corresponding to the 3D models with $\Delta t = 5 \times 10^{-5}$ and $\Delta x$ (=$\Delta y$=$\Delta z$) = 0.1 km. Sintering commences around 0.25 Myr.

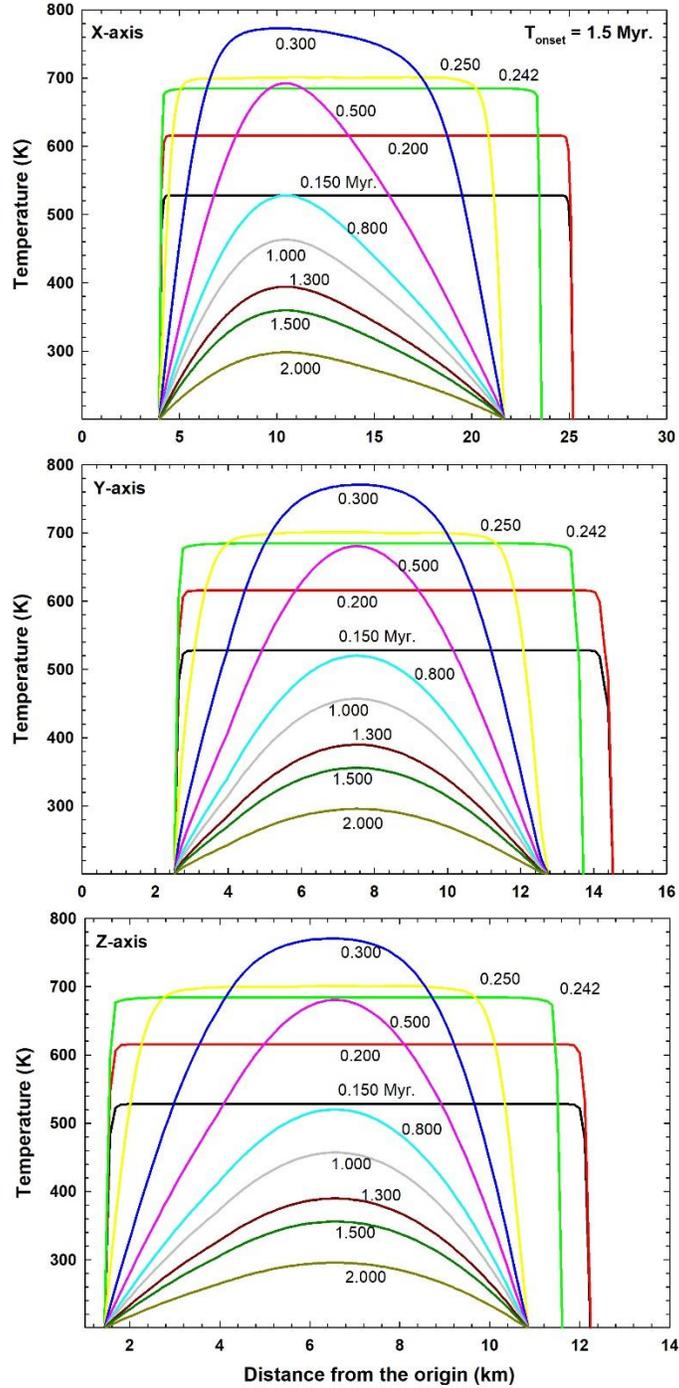

**Figure 2.** Thermal profiles of (951) Gaspra alike asteroid (fig. 1). A $T_{Onset}$ time of 1.5 Myr was assumed for the onset of instantaneous accretion. The explicit 3D-FDM (finite difference method) was used to deduce the thermal profiles of the body with an initial temperature of 200 K and an identical constant surface temperature. All the timescales are defined with respect to the onset time of the accretion of the asteroid. All distances are measured with respect to the origin of the coordinate system.

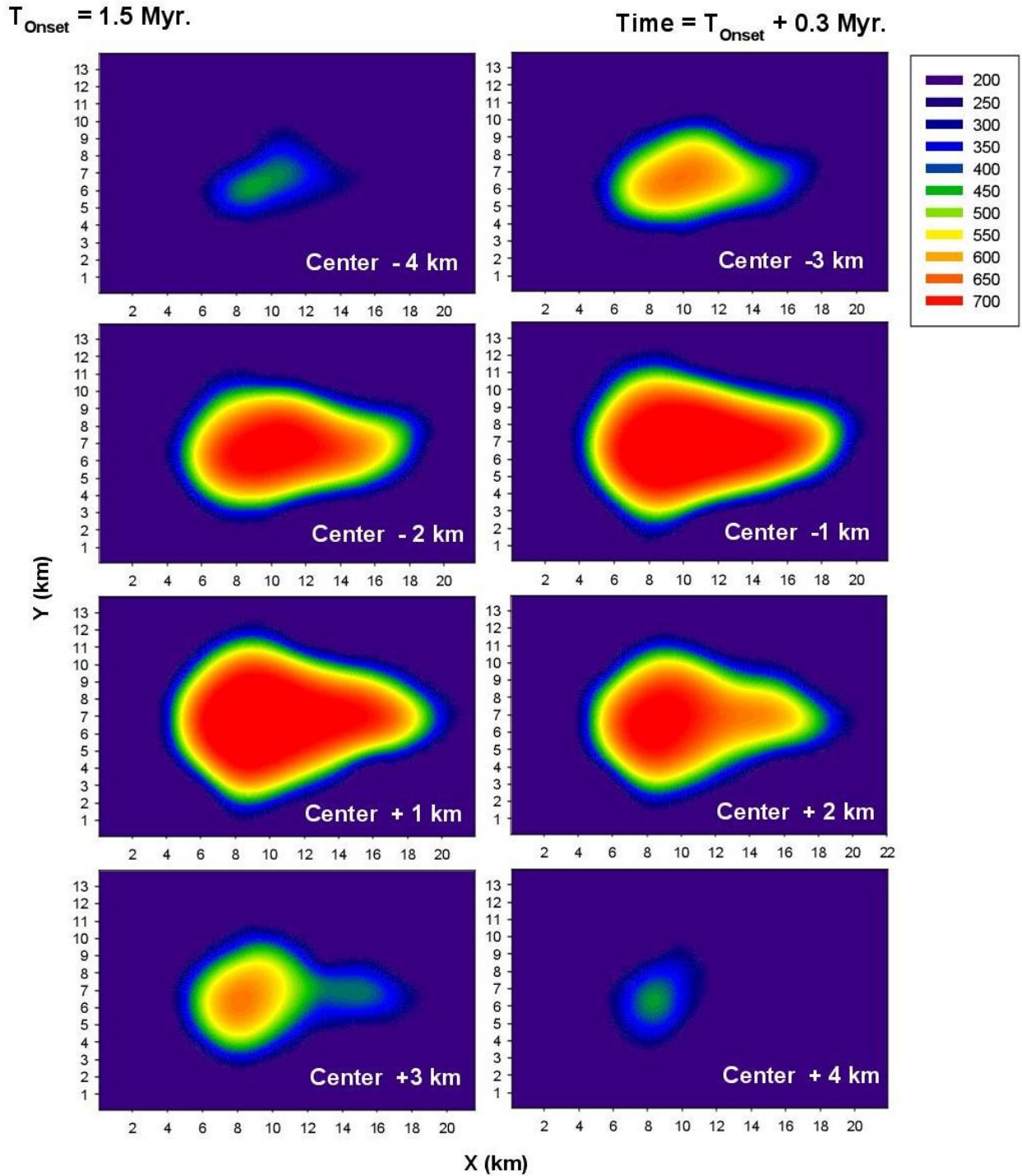

**Figure 3.** Temperature (K) contours along the eight XY planes of an (951) Gaspra alike asteroid at 0.3 Myr. subsequent to accretion at $T_{Onset}$ = 1.5 Myr. These eight planes are orthogonal to Z-axis at a distance of -4, -3, -2, -1, 1, 2, 3 and 4 km with respect to the center of the Z-axis.

**Figure 4.** Temperature (K) contours of (951) Gaspra alike asteroid with $T_{Onset}$ = 3.5 Myr. Temperature contours for the three planes, XY, XZ and YZ through the center of the body at four distinct time corresponding to the 3D models with $\Delta t = 5 \times 10^{-4}$ and $\Delta x\ (=\Delta y=\Delta z) = 0.1$ km. Sintering does not occur in this scenario.

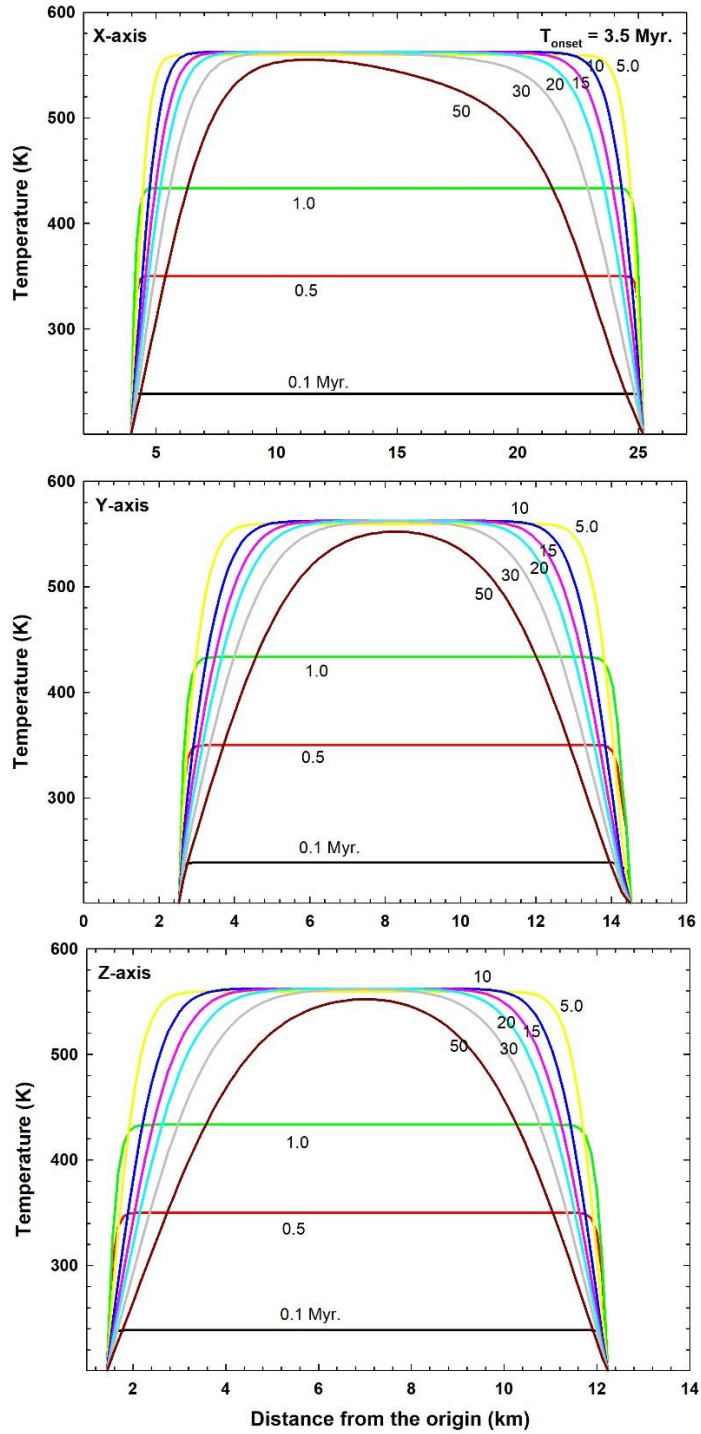

**Figure 5.** Thermal profiles of (951) Gaspra alike asteroid (fig. 3). A $T_{Onset}$ time of 3.5 Myr was assumed for the onset of instantaneous accretion.

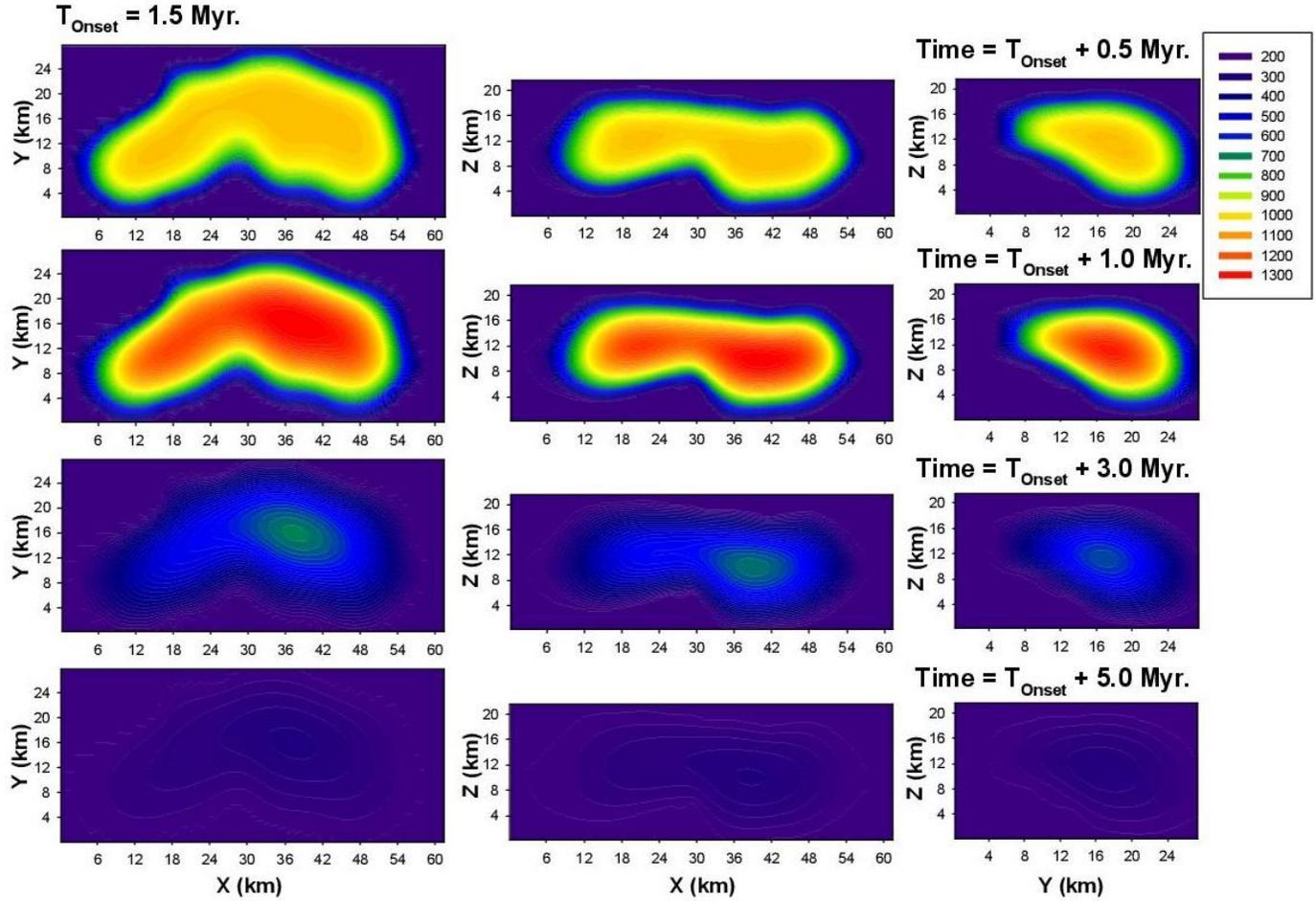

**Figure 6.** Temperature (K) contours of (243) Ida alike asteroid with $T_{Onset}$ = 1.5 Myr. Temperature contours for the three planes, XY, XZ and YZ through the center of the body at four distinct time corresponding to the 3D models with $\Delta t = 1 \times 10^{-4}$ and $\Delta x$ (=$\Delta y$=$\Delta z$) = 0.2 km. Sintering occurs around 0.25 Myr.

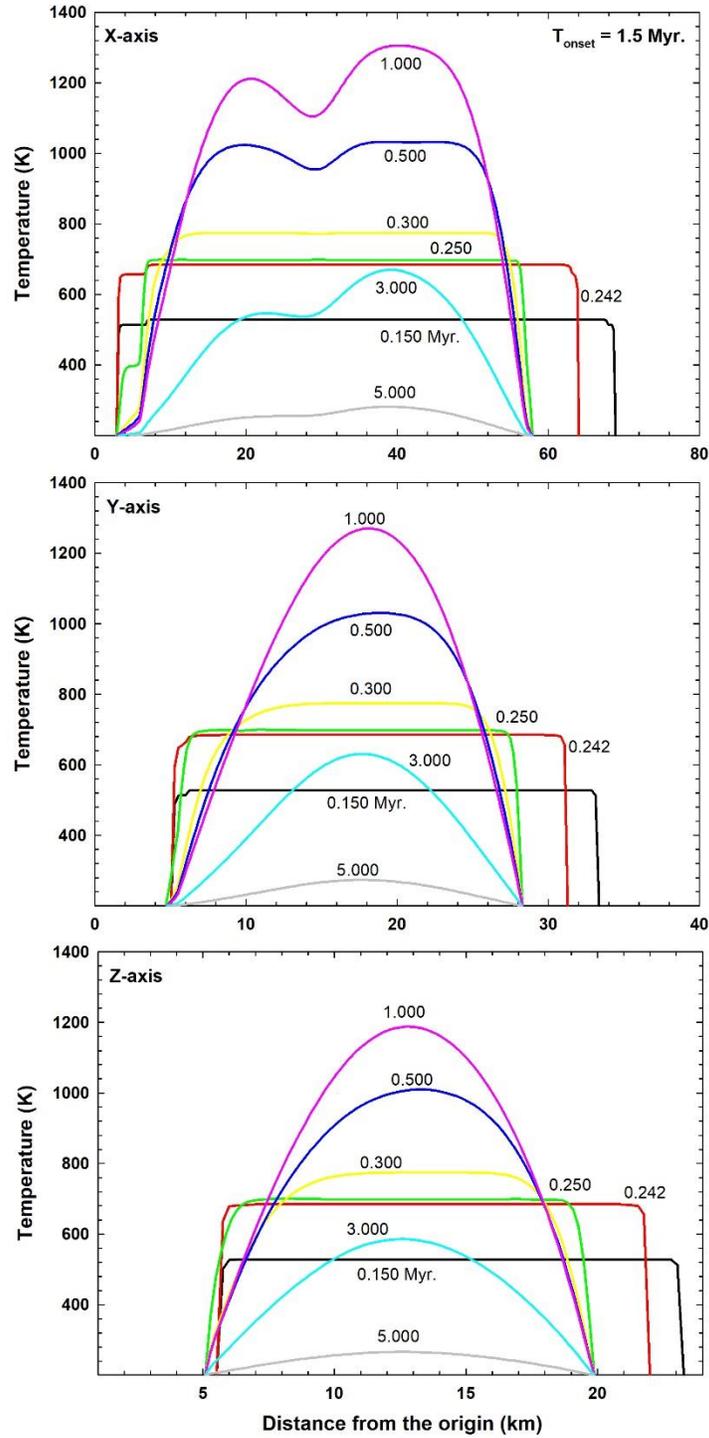

**Figure 7.** Thermal profiles of (243) Ida alike asteroid (fig. 5). A T$_{Onset}$ time of 1.5 Myr was assumed for the onset of instantaneous accretion.